\documentclass[12pt]{article}
\usepackage[english,activeacute]{babel}
\usepackage{natbib}
\usepackage{comment}
\usepackage{float}
\usepackage[hidelinks]{hyperref}
\usepackage{mathrsfs}
\usepackage{enumitem}
\usepackage[font={small,it}]{caption}
\usepackage{amsmath,amsfonts,amsthm,amssymb}
\usepackage{bm,rotating,multirow,dsfont,graphicx}
\usepackage[usenames, dvipsnames]{color}
\usepackage{url}
\usepackage{multicol}
\usepackage{multirow}
\usepackage[T1]{fontenc}
\usepackage{flafter}
\usepackage{appendix}
\usepackage{subfigure}
\usepackage{xcolor}
\makeatletter
\def\hlinewd#1{%
	\noalign{\ifnum0=`}\fi\hrule \@height #1 %
	\futurelet\reserved@a\@xhline}
\makeatother
\def\@roman#1{\romannumeral #1}

\begin{document}

\def\spacingset#1{\renewcommand{\baselinestretch}{#1}\small\normalsize}\spacingset{1}

\title{Conditional Relative Risk:\\ An association measure for longitudinal data analysis}

\date{}

\author{
	Lina Buitrago\footnote{labuitragor@unal.edu.co}, Juan Sosa\footnote{jcsosam@unal.edu.co}, Oscar Melo\footnote{oomelom@unal.edu.co}  \\
	Department of Statistics, Faculty of Sciences\\
 Universidad Nacional de Colombia, Bogota, Colombia
}	 
	       
\maketitle

\begin{abstract} 
In this paper, we propose a novel association measure for longitudinal studies based on the traditional definition of relative risk. In a Markovian fashion, such a proposal takes into account the information content regarding the previous time. We derive its corresponding confidence interval by means of the Delta method having in mind the crude association between factor and event. Also, we study the properties of our uncertainty quantification scheme through an exhaustive simulation study. Our findings show that the coverage probability is quite close to the level of confidence. Finally, our proposal has a reasonable interpretation from the epidemiological as well as the statistical point of view.
\end{abstract}


\spacingset{1.1} % DON'T change the spacing!

\section{Introduction}

In many longitudinal studies, it is necessary to measure the outcome of interest on several occasions throughout the follow-up period \citep{Twisk05}. In these cases, traditional crude association measures as well as those adjusted through linear models \citep{Rothman19} are typically not appropriate, since they do not take into account the existing correlation among measurements corresponding to the same individual.

Traditionally, either cohort or experimental longitudinal studies with dichotomous outcomes use the relative risk (RR) or the odds ratio (OR) as measures of association. These measures can be calculated either adjusted or crude, depending on whether or not confounding variables are taken into account \citep{Rothman19,Szklo19}. On one side, when considering crude association, confidence intervals are typically obtained through a direct application of the Delta method \citep{Morris89}. However, when controlling for confounding variables is in order, confidence intervals are based on linear model fitting. 

Models for longitudinal data are commonly classified into marginal models, transition models and random effects models \citep{Diggle02}. To date, there are available several techniques to provide estimates for parameters in marginal models mainly relying on the generalized estimation equations (EEG, \citealt{Liang86}). These methods have been extended in several directions, such as those provided by \citep{Lipsitz91}, who modified the equations in order to find an estimate based on the OR, and also, considering not fixed by design follow-up times with outcomes that may depend on previous outcomes \citep{Fitzm06}. 
Other types of marginal models have also been developed, such as structural marginal models, which are closely linked to causal inference approaches in Epidemiology (e.g., \citealt{Robins00}). 

In model-based approaches, correlation effects are implicitly included when estimating model parameters through a user-prespecified covariance matrix, but not explicitly when estimating association parameters such as the RR or the OR. 
Thus, we propose a way to include directly the correlation between the occurrence of the event in two moments $k$ and $j$, with $k < j$. To do so, we develop correlated binary data theory in order to develop two brand new association measures based on the RR. Specifically, we propose two crude conditional association measures for $2\times 2$ tables, including point estimates as well as uncertainty quantification (confidence intervals).

This paper is structured as follows: Section \ref{sec2} develops theory for correlated binary data, Section \ref{sec3} derives novel measures of association for $2\times 2$ tables along with the corresponding confidence intervals, Section \ref{sec4} makes a comparison between our proposal and traditional RR, Section \ref{sec5} shows an actual application with experimental data, and finally, Section \ref{sec6} discusses findings and recommendations for future research.

\section{Theory for correlated binary data}\label{sec2}

An association measure for longitudinal studies is the relative risk $RR$, given by
\begin{equation}
RR=\frac{\pi_E}{\pi_{NE}} \label{rr}
\end{equation}
where $\pi_E$ is the outcome risk between exposed and $\pi_{NE}$ is the outcome risk between non-exposed.

In this case, we want to establish an association measure between a dichotomous exposure with a dichotomous outcome at time $j$, conditioned by the outcome at time $k$, with $k < j$. 
The conditional probability of occurrence of the outcome at $j$ is calculated below for two cases, namely, assuming that the outcome's probability remains constant over the follow-up period, and also, assuming that this probability varies over the follow-up period.

\subsection{The outcome's probability remains constant over the follow-up period}

Let $\boldsymbol{Y}=(Y_{1},Y_{2},\ldots,Y_{t})$ be the random vector with the response variable over the follow-up period, with $Y_{j}\sim \textsf{Bernoulli}(\pi)$, where $\pi$ is the probability of occurrence of the outcome at time $j$:
\begin{equation*}
	Y_j=
        \begin{cases}
	1 & \text{ if the outcome happens at time $j$}\\
	0 &\text{ if the outcome does not happens at time $j$.}\\
	\end{cases}
\end{equation*}
Thus, 
\begin{equation}
    \textsf{E}\left(Y_{j}\right)=\pi\,,\qquad
    \textsf{Var}\left(Y_{j}\right)=\pi(1-\pi)\,,\qquad
    \textsf{Cov}\left(Y_{j},Y_{k}\right)=\rho_{jk}\pi(1-\pi)\,,\label{covy}
\end{equation}				
where $\rho_{jk}$ is the correlation between the outcomes at times $j$ and $k$. Additionally, since
$\textsf{Cov}\left(Y_{j},Y_{k}\right)=\textsf{E}\left(Y_{j}Y_{k}\right)-\textsf{E}\left(Y_{j}\right)\textsf{E}\left(Y_{k}\right)$  	
and
\begin{align}\label{pcon}
    \textsf{E}\left(Y_{j}Y_{k}\right)=\sum_{y_j,y_k}{y_{j}y_{k}\Pr(Y_{j}=y_{j},Y_{k}}=y_{k})
                            =\Pr(Y_{j}=1,Y_{k}=1)\,,
\end{align} 
Then, from equations \eqref{covy} and \eqref{pcon}, it follows that 
\begin{align*}
    \textsf{Cov}\left(Y_{j},Y_{k}\right) = \Pr\left(Y_{j}=1,Y_{k}=1\right)-{\pi}^2 = \rho_{jk}\pi(1-\pi)\,.
\end{align*}

Now, in the same spirit of \cite{Crowder85}, we obtain that
\begin{align*}
    \Pr\left(Y_{j}=1\mid Y_{k}=1\right)=\frac{\Pr\left(Y_{j}=1,Y_{k}=1\right)}{\Pr\left(Y_{k}=1\right)} = \pi+\rho_{jk}(1-\pi)\,,
\end{align*}
and similarly,
\begin{equation*}
\Pr\left(Y_{j}=1\mid Y_{k}=0\right)=(1-\rho_{jk})\pi\,.
\end{equation*}

Finally, since both $0\leq\Pr\left(Y_{j}=1\mid Y_{k}=1\right)\leq 1$ and $0\leq\Pr\left(Y_{j}=1\mid Y_{k}=0\right)\leq 1$, then the correlation between the outcomes at times $j$ and $k$ is bounded in a way that
\begin{equation*}
    \max\left\lbrace-\frac{\pi}{1-\pi},-\frac{1-\pi}{\pi}\right\rbrace\leq\rho_{jk}\leq 1\,.
\end{equation*}

\subsection{The outcome's probability changes over the follow-up period}

As of now, we have made the assumption that the outcome's probability remains constant over the follow-up period, but usually the individual's conditions change over time, thus the outcome's probability changes too. So, now we let $Y_{j}\sim \textsf{Bernoulli}(\pi_{j})$, where $\pi_{j}$ is the outcome's probability at time $j$.

Under these conditions, we get that
\begin{equation*}
    \textsf{E}\left(Y_{j}\right)=\pi_j\,,\qquad
    \textsf{V}\left(Y_{j}\right)=\pi_j(1-\pi_j)\,,\qquad
    \textsf{Cov}\left(Y_{j},Y_{k}\right)=\rho_{jk}\sqrt{\pi_j(1-\pi_j)\pi_k(1-\pi_k)}\,.
\end{equation*}
Following the same reasoning as before, it follows that
\begin{equation}
    \Pr\left(Y_{j}=1\mid Y_{k}=1\right)=\pi_{j}+\ \rho_{jk}\sqrt{\frac{\pi_{j}(1-\pi_{j})(1-\pi_{k})}{\pi_{k}}}
\end{equation} \label{pcond1kj}
as well as
\begin{equation}
    \Pr\left(Y_{j}=1\mid Y_{k}=0\right)=\pi_{j}-\rho_{jk}\sqrt{\frac{\pi_{j}(1-\pi_{j})\pi_{k}}{1-\pi_{k}}}\,.
\end{equation}\label{pcond0kj}

Once again, since both
$0\leq\Pr\left(Y_{j}=1\mid Y_{k}=1\right)\leq 1$ and $0\leq\Pr\left(Y_{j}=1\mid Y_{k}=0\right)\leq 1$, then the correlation between the outcomes at times $j$ and $k$ is bounded in a way that
\begin{equation*}
    -\sqrt{\frac{\pi_{j}\pi_{k}}{\left(1-\pi_{j}\right)\left(1-\pi_{k}\right)}}\leq\rho_{jk}\leq\sqrt{\frac{\pi_{k}\left(1-\pi_{j}\right)}{\pi_{j}\left(1-\pi_{k}\right)}}\,,
\end{equation*}
and also,
\begin{equation*}
    -\sqrt{\frac{\left(1-\pi_{j}\right)\left(1-\pi_{k}\right)}{\pi_{j}\pi_{k}}}\leq\rho_{jk}\leq\sqrt{\frac{\left(1-\pi_{j}\right)\left(1-\pi_{k}\right)}{\pi_{j}\pi_{k}}}\,.
\end{equation*}

\section{A new association measure}\label{sec3}

Taking as starting point the classical definition of relative risk \eqref{rr} at time $j$, we are able to include the information about what actually happened at time $k$, with $k<j$. Using the result in \eqref{pcond1kj}, the relative risk given that the outcome occurred at $k$ is defined as:
\begin{align} \label{rr1}
{RR}_1&=\frac{\pi_{Ej\mid k=1}}{\pi_{\bar{E}j\mid k=1}}=\frac{{\Pr\left(Y_{j}=1\mid Y_{k}=1\right)}_E}{{\Pr\left(Y_{j}=1\mid Y_{k}=1\right)}_{\bar{E}}}\\\notag
    &=\frac{\pi_{jE}+\rho_{jkE}\sqrt{\pi_{jE}(1-\pi_{jE})(1-\pi_{kE})/\pi_{kE}} }{\pi_{j\bar{E}}+\rho_{jk\bar{E}}\sqrt{\pi_{j\bar{E}}(1-\pi_{ij\bar{E}})(1-\pi_{k\bar{E}})/\pi_{k\bar{E}}}}\,, 
\end{align}
where $\pi_{Ej|k=1}$ is the outcome's probability at time $j$ given that the outcome occurred at $k$ among the exposed, $\pi_{\bar{E}j|k=1}$ is the outcome's probability at time $j$ given that the outcome occurred at $k$ among the non-exposed, $\pi_{jE}$ is the outcome's probability at time $j$ among the exposed, $\pi_{j\bar{E}}$ is the outcome's probability at time $j$ among the non-exposed, $\rho_{jkE}$ is the correlation between the dichotomous variables $Y_j$ and $Y_k$ for the exposed, and finally, $\rho_{jk\bar{E}}$ is the correlation for the dichotomous variables $Y_j$ and $Y_k$ for the non-exposed.
Additionally, we define the relative risk at time $k$ given that the outcome did not occur at time $j$ as:
\begin{align}
    {RR}_0&=\frac{\pi_{Ej|k=0}}{\pi_{\bar{E}j|k=0}}=\frac{{\Pr\left(Y_{j}=1|Y_{k}=0\right)}_E}{{\Pr\left(Y_{j}=1|Y_{k}=0\right)}_{\bar{E}}}\\\notag
    &=\frac{\pi_{jE}-\rho_{jkE}\sqrt{\pi_{jE}(1-\pi_{jE})\pi_{kE}/(1-\pi_{kE})}}{\pi_{j\bar{E}}-\rho_{jk\bar{E}}\sqrt{\pi_{j\bar{E}}(1-\pi_{j\bar{E}})(\pi_{k\bar{E}})/(1-\pi_{k\bar{E}})}}\,,\label{rr0}
\end{align}
where $\pi_{Ej|k=0}$, $\pi_{\bar{E}j|k=0}$, $\pi_{jE}$, $\rho_{jkE}$, and $\rho_{jk\bar{E}}$ are defined in a analogous manner as above. 

\subsection{Inference}

Typically, it is impossible to work directly with the entire population, and as a consequence, we need to estimate the population's parameters $RR_1$ and $RR_0$. In this spirit, we can see the problem as a problem with two tables, namely, one table for the sample individuals for whom the outcome occurred at time $k$ (Table \ref{table1}), and another table for the sample individuals for whom the outcome did not occur at time $k$ (Table \ref{table2}), with $k<j$.
\begin{table}[!ht]
    \caption{\small{Sampling table at time $j$ given that the outcome occurred at time $k$. }}\label{table1}
\centering {\small
\begin{tabular}{cccc}\hline
   &
Outcome (Yes) &
Outcome (Not)   &
Total \\ \hline
Exposed (Yes)    & $a_1$  &  $b_1$  & $a_1+b_1=n_{1E}$  \\
Exposed (Not)    & $c_1$  &  $d_1$  & $c_1+d_1=n_{1\bar{E}}$  \\ \hline
\end{tabular}}
\end{table} 

\begin{table}[!ht]
\caption{\small{Sampling table at time $j$ given that the outcome did not occur at time $k$. }}\label{table2}
\centering {\small
\begin{tabular}{cccc}\hline
   &
Outcome (Yes) &
Outcome (Not)   &
Total \\ \hline
Exposed (Yes)    & $a_0$  &  $b_0$  & $a_0+b_0=n_{0E}$  \\
Exposed (Not)    & $c_0$  &  $d_0$  & $c_0+d_0=n_{0\bar{E}}$  \\ \hline
\end{tabular}}
\end{table} 

This way, the point estimates for $RR_1$ and $RR_0$ are given by:
\begin{equation}\label{rr1e}
\widehat{RR}_1 =\frac{\hat{\pi}_{Ej|k=1}}{\hat{\pi}_{\bar{E}j|k=1}}=\frac{a_1/n_{1E}}{c_1/n_{1\bar{E}}}
    =\frac{\hat{\pi}_{jE}+\hat{\rho}_{jkE}\sqrt{\hat{\pi}_{jE}(1-\hat{\pi}_{jE})(1-\hat{\pi}_{kE})/\hat{\pi}_{kE}}\ }{\hat{\pi}_{j\bar{E}}+\hat{\rho}_{jk\bar{E}}\sqrt{\hat{\pi}_{j\bar{E}}(1-\hat{\pi}_{ij\bar{E}})(1-\hat{\pi}_{k\bar{E}})/\hat{\pi}_{k\bar{E}}}} 
\end{equation}
and
\begin{equation}\label{rr0e}
\widehat{RR}_0=\frac{\hat{\pi}_{Ej|k=0}}{\hat{\pi}_{\bar{E}j|k=0}}=\frac{a_0/n_{0E}}{c_0/n_{0\bar{E}}}
    =\frac{\hat{\pi}_{jE}-\hat{\rho}_{jkE}\sqrt{\hat{\pi}_{jE}(1-\hat{\pi}_{jE})\hat{\pi}_{kE}/(1-\hat{\pi}_{kE})}\ }{\hat{\pi}_{j\bar{E}}-\hat{\rho}_{jk\bar{E}}\sqrt{\hat{\pi}_{j\bar{E}}(1-\hat{\pi}_{j\bar{E}})(\hat{\pi}_{k\bar{E}})/(1-\hat{\pi}_{k\bar{E}})}}\,.
\end{equation}
Furthermore, it is also possible to estimate the correlations $\rho_{jkE}$ and $\rho_{jk\bar{E}}$ as follows:
\begin{equation*}
    {\hat{\rho}}_{ijkE}=\frac{a_1n_{0E}-a_0n_{1E}}{\sqrt{n_{1E}n_{0E}(a_1+a_0)(b_1+b_0)}}
    \quad\text{and}\quad
    {\hat{\rho}}_{ijk\bar{E}}=\frac{c_1n_{0\bar{E}}-c_0n_{1\bar{E}}}{\sqrt{n_{1\bar{E}}n_{0\bar{E}}(c_1+c_0)(b_1+b_0)}}\,.
\end{equation*}

\subsubsection{Confidence interval}

Under the frequentist paradigm, our proposal is based on the probabilistic distribution of $\ln{RR_1}$ and $\ln{RR_0}$ for a large sample sizes $n_{1E}$ and $n_{0E}$. By the central limit theorem, we have that
\begin{equation*}
\sqrt{n_{1E}}\left(\hat{\pi}_{Ej|k=1}-\pi_{Ej|k=1}\right)\xrightarrow{d} \textsf{N}\left(0, \pi_{Ej|k=1}(1-\pi_{Ej|k=1})\right)\,,
\end{equation*}
and therefore, by taking logarithms and applying the Delta method, we finally get that
\begin{equation*}
    \sqrt{n_{1E}}\left(\ln\hat{\pi}_{Ej|k=1}-\ln\pi_{Ej|k=1}\right)\xrightarrow{d} \textsf{N}\left(0,\frac{1-\pi_{Ej|k=1}}{\pi_{Ej|k=1}}\right)\,.
\end{equation*}
Since $\ln{\widehat{RR}_1}=\ln\hat{\pi}_{Ej|k=1}-\ln\hat{\pi}_{j\bar{E}}$, it follows that an asymptotic confidence interval for $\ln{RR_1}$ with $100(1-\alpha)\%$ confidence is given by
\begin{equation}\label{intlnRR1}
    \ln{\widehat{RR}_1}\pm z_{1-\alpha/2}\sqrt{\frac{1-\hat\pi_{Ej|k=1}}{n_{1E}\hat\pi_{Ej|k=1}}+\frac{1-\hat\pi_{\bar{E}j|k=1}}{n_{1\bar{E}}\hat\pi_{\bar{E}j|k=1}}}\,,
\end{equation}
where $z_{1-\alpha/2}$ is the percentile $100(1-\alpha/2)$ of the standard normal distribution. Similarly, 
it follows that an asymptotic confidence interval for $\ln{RR_0}$ with $100(1-\alpha)\%$ confidence is given by
\begin{equation}\label{intlnRR0}
    \ln{\hat{RR}_0}\pm z_{1-\alpha/2}\sqrt{\frac{1-\hat\pi_{Ej|k=0}}{n_{0E}\hat\pi_{Ej|k=0}}+\frac{1-\hat\pi_{\bar{E}j|k=0}}{n_{0\bar{E}}\hat\pi_{\bar{E}j|k=0}}}\,.
\end{equation}
Taking exponential in \eqref{intlnRR1} and \eqref{intlnRR0}, the corresponding confidence intervals for $RR_1$ and $RR_0$ with $100(1-\alpha)\%$ confidence are respectively:
\begin{equation}
    \hat{RR}_1\exp{\left(\pm z_{1-\alpha/2}\sqrt{\frac{1-\hat\pi_{Ej|k=1}}{n_{1E}\hat\pi_{Ej|k=1}}+\frac{1-\hat\pi_{\bar{E}j|k=1}}{n_{1\bar{E}}\hat\pi_{\bar{E}j|k=1}}}\right)}\label{intRR1}
\end{equation}
and
\begin{equation}
    \hat{RR}_0\exp{\left(\pm z_{1-\alpha/2}\sqrt{\frac{1-\hat\pi_{Ej|k=0}}{n_{0E}\hat\pi_{Ej|k=0}}+\frac{1-\hat\pi_{\bar{E}j|k=0}}{n_{0\bar{E}}\hat\pi_{\bar{E}j|k=0}}}\right)}\,.\label{intRR0}
\end{equation}

\subsubsection{Coverage probability}

The coverage probability $P_c$ of a confidence interval is the probability that the interval contains the true parameter value. Then, an optimal result occurs when the coverage probability is near to the corresponding confidence, i.e., $Pc\approx 1-\alpha$, or alternatively, $1-Pc\approx\alpha$.
In order to compute $P_c$ for $RR_1$ and $RR_0$ considering $100(1-\alpha)\%=95\%$ confidence, we simulate 2025 ($3\times 3\times 5 \times 5 \times 3 \times 3$) scenarios according to the following cases: 
\begin{itemize}
    \item Sample sizes $n_E$ and $n_{\bar{E}}$ vary on $\lbrace500, 1000, 2000\rbrace$ ($3\times 3$ possibilities). 
    \item Occurrence probabilities $\pi_{ijE}$ and $\pi_{ij\bar{E}}$ are considered constant over the follow up period and vary on $\lbrace0.1,0.3,0.5,0.7,0.9\rbrace$ ($5\times 5$ possibilities).
    \item Correlations between exposed $\rho_{ijkE}$ and non-exposed $\rho_{ijk\bar{E}}$ are considered as low  medium, and high correlation and vary on $\lbrace0.1,0.5,0.9\rbrace$ ($3\times 3$ possibilities).
\end{itemize}

We use the following procedure to obtain the corresponding coverage probability en each of the 2025 scenarios:
\begin{enumerate}
    \item Generate every $2\times2$ table with entries greater than 0 given that both the outcome occurred and the outcome did not occur at time $k$, with $k < j$. 
    \item For each table, compute the 95\% confidence interval for ${RR}_k$. 
    \item Identify if the true value of ${RR}_k$ lies within the corresponding interval. 
   \item Compute $P_c$ as
    \begin{equation}
        P_c=\sum_{t\in T}p_t\,,
    \end{equation}
    where $p_t$ is the probability of occurrence of table $t$ and $T$ is the set of those tables whose corresponding interval covers the true value of ${RR}_k$. 
\end{enumerate}

\subsubsection{Results}

Recall that optimal results are obtained when $1-P_c\approx\alpha$. We obtain in all cases that $1-P_c < 0.09$. Additionally, for the $RR_1$ confidence interval:

\begin{enumerate}
    \item When the outcome's probability for non-exposed is $\pi_{j\bar{E}}=0.1$ (see Figure \ref{figure1}):
    \begin{itemize}
        \item For $\rho_{jk\bar{E}}=0.1$, $1-P_c$ is lower when $\pi_{jE}$ is very low or very high. Moreover, $1-P_c$ is close to $\alpha$ when $\pi_{jE}$ assumes intermediate values.
        \item For $\rho_{jk\bar{E}}=0.5$, in general, $1-P_c$ is greater than in the previous case. As $\pi_{jE}$ increases, $1-P_c$ slowly increases.
        \item For $\rho_{jk\bar{E}}=0.9$, $1-P_c$ takes the highest values of them all. As $\pi_{jE}$ increases, $1-P_c$ increases up to values close to $0.09$ approximately, except when $\rho_ {jkE} = 0.9$, where $1-P_c$ increases for small values of $\pi_{jE}$, and then, decreases for large values of $\pi_{jE}$.
    \end{itemize}
\begin{figure}[!ht]
  \centering
  \includegraphics[scale=0.95]{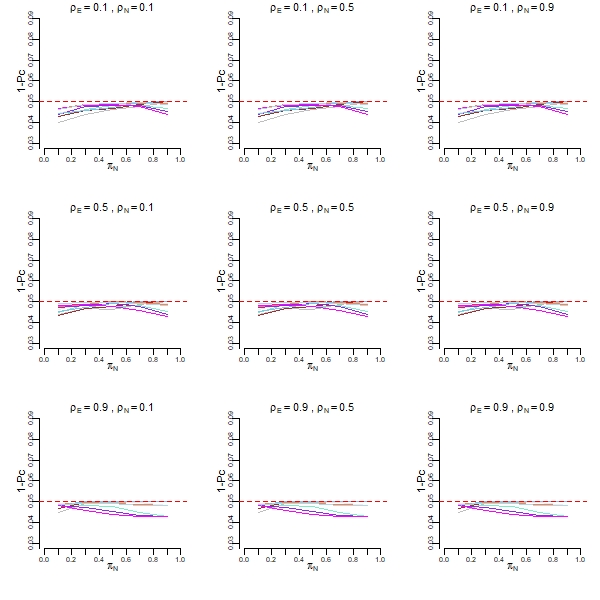}
  \caption{Complement of the coverage probability of $RR_1$ confidence intervals when $\pi_{j\bar{E}}=0.1$.}\label{figure1}
\end{figure}

\item When the outcome's probability for non-exposed is $\pi_{j\bar{E}}=0.5$ (see Figure \ref{figure2}):
    \begin{itemize}
        \item For $\rho_{jkE}=0.1$, $1-P_c<0.05$ when $\pi_ {ijE}$ assumes low values. As $\pi_ {ijE}$ increases, $1-P_c$ is close to $\alpha$ and remains almost constant until the highest values of $\pi_{ijE}$ are reached, case in which $1-P_c$ is greater.
        \item For $\rho_{jkE}=0.5$, in general, $1-P_c$ is greater than in the previous case.
        \item For $\rho_{jkE}=0.9$, $1-P_c$ takes the highest values of them all. When $\pi_{jE}=0.1$, $1-P_c$ decreases down to values close to $0.05$, unless $\rho_{jk\bar{E}}=0.9$, where $1-P_c$ decreases for intermediate values of $\pi_{jE}$, and then, it increases to higher values of $\pi_{jE}$.
    \end{itemize}
\begin{figure}[!ht]
  \centering
  \includegraphics[scale=0.95]{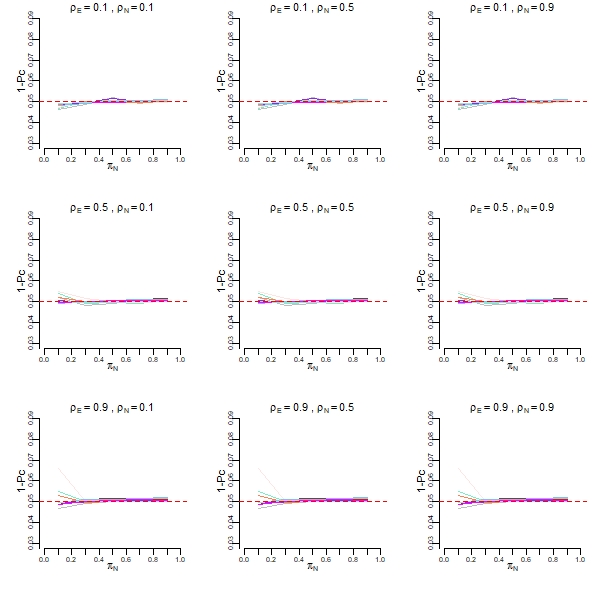}
  \caption{Complement of the coverage probability of $RR_1$ confidence intervals when $\pi_{j\bar{E}}=0.5$.}\label{figure2}
\end{figure}

\item When the outcome's probability for non-exposed is $\pi_{j\bar{E}}=0.9$ (see Figure \ref{figure3}):
    \begin{itemize}
        \item Results are very similar to those obtained in the previous instance, but we can see more variability in the results of $1-P_c$, depending on the values of ${n}_E$ and $n_{\bar{E}}$.
        \item Typically, $1-P_c$ assumes high values, especially when $\rho_{ijkE}=0.9$.
        \item Also, $1-P_c$ has a substantial decrease for intermediate values of $\pi_{ijE}$, with a slight increase for highest values of $\pi_{ijE}$.
    \end{itemize}
\begin{figure}[!ht]
  \centering
  \includegraphics[scale=0.95]{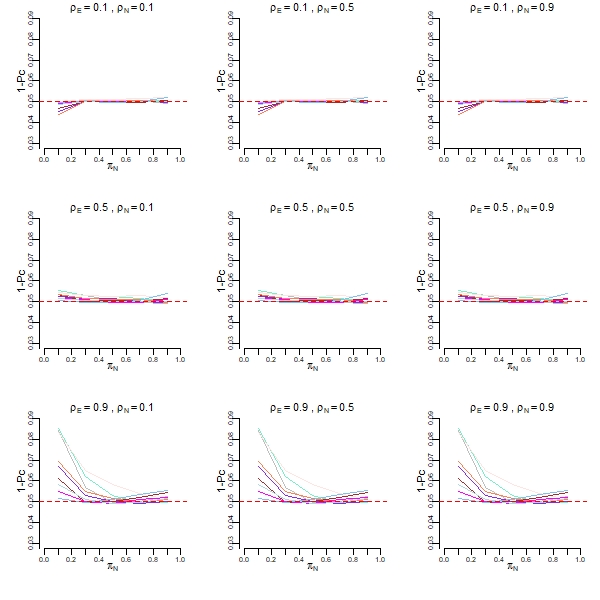}
  \caption{Complement of the coverage probability of $RR_1$ confidence intervals when $\pi_{j\bar{E}}=0.9$.}\label{figure3}
\end{figure}

\end{enumerate}

In general, with respect to samples sizes, the largest sample sizes of non-exposed ($n_{\bar{E}}$) are those that produce the values of $1-P_c$ close to $0.05$ and vice-versa, so that the closest value to $0.05$ is obtained when $({n} _E, n\bar{E})=(2000,2000)$.

\begin{figure}[!ht]
  \centering
  \includegraphics[scale=0.95]{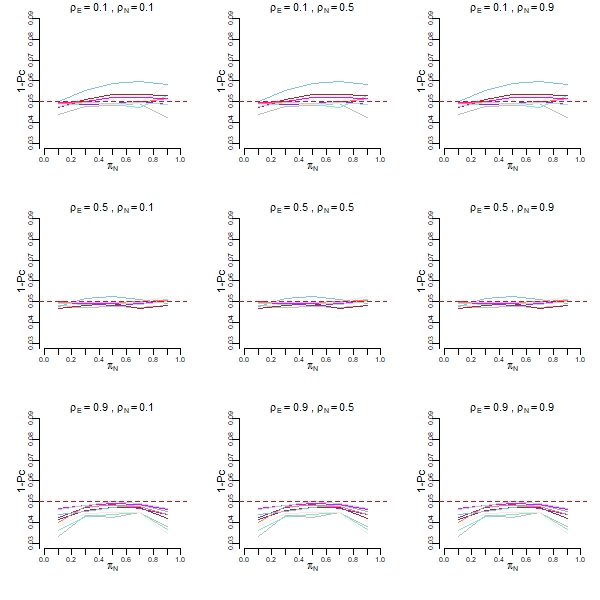}
  \caption{Complement of the coverage probability of $RR_0$ confidence intervals when $\pi_{j\bar{E}}=0.1$.}\label{figure4}
\end{figure}

\begin{figure}[!ht]
  \centering
  \includegraphics[scale=0.95]{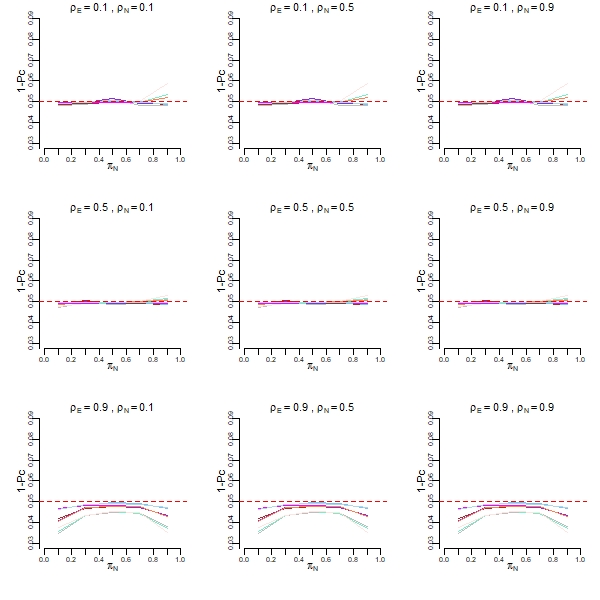}
  \caption{Complement of the coverage probability of $RR_0$ confidence intervals when $\pi_{j\bar{E}}=0.5$.}\label{figure5}
\end{figure}

\begin{figure}[!ht]
  \centering
  \includegraphics[scale=0.95]{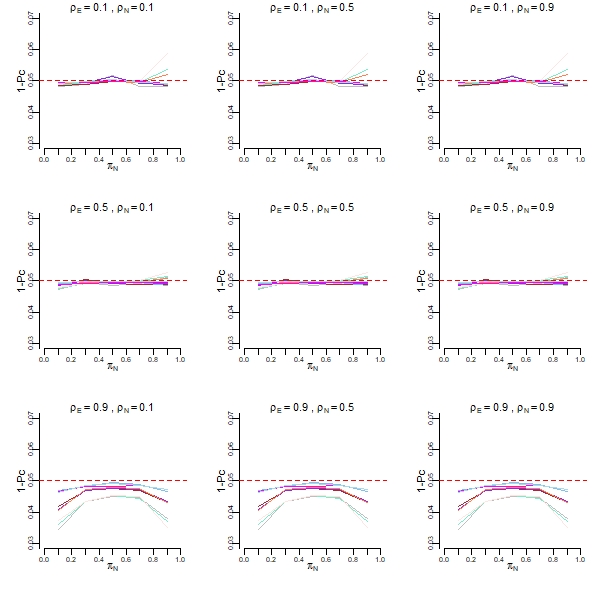}
  \caption{Complement of the coverage probability of $RR_0$ confidence intervals when $\pi_{j\bar{E}}=0.9$.}\label{figure6}
\end{figure}

Finally, results for $RR_0$ can be interpreted in a similar fashion as above. See Figures \ref{figure4}, \ref{figure5}, and \ref{figure6} for details.

\section{Comparison with conventional relative risk measures} \label{sec4}

In order to evidence the effect of considering the correlation with what happened in the previous moment, we compare conventional relative risk measures with our proposal.

\subsection{$RR_1$ versus \bf$RR$}

\begin{figure}[!ht]
  \centering
  \includegraphics [scale=0.6]{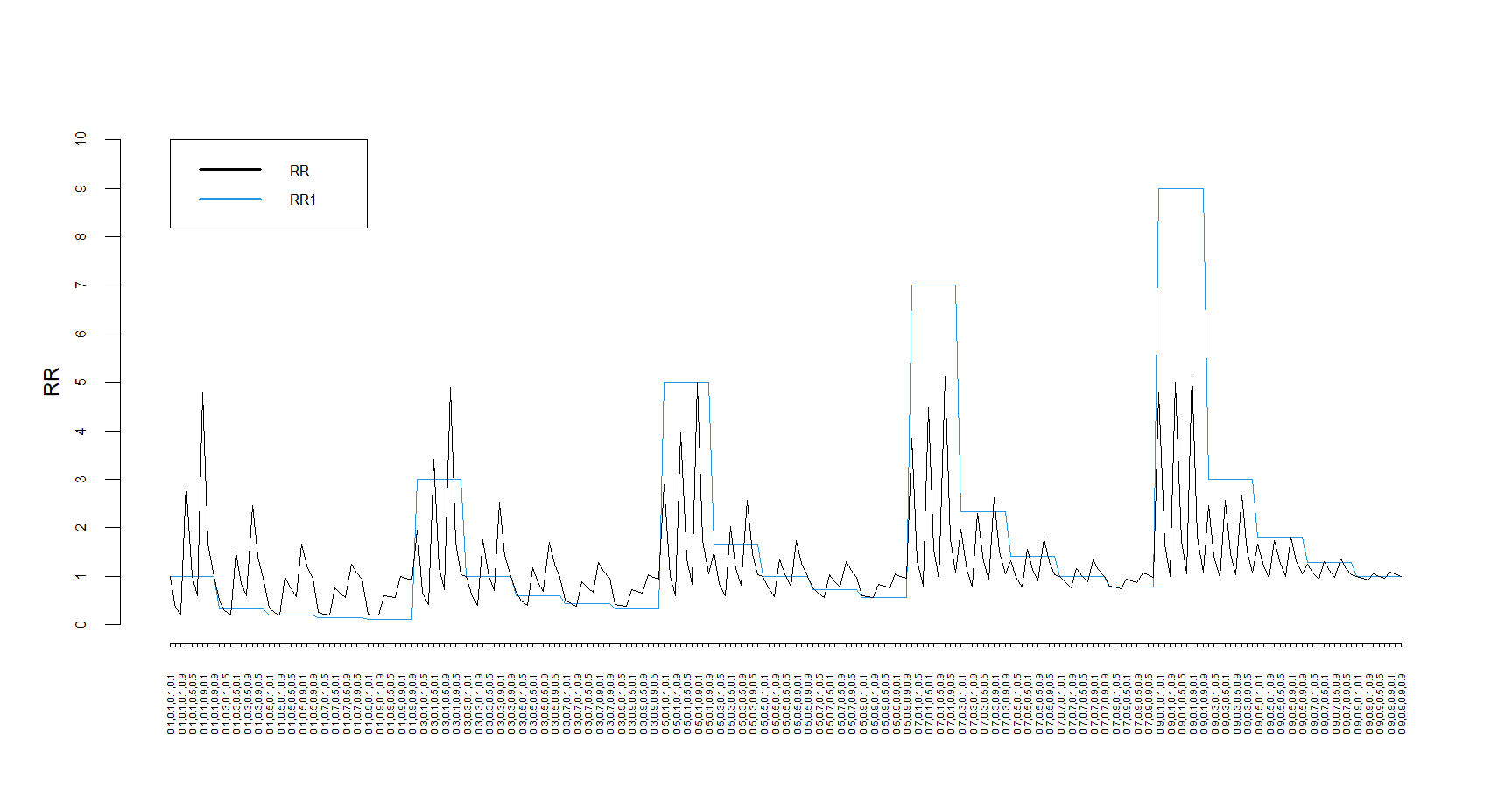}
  \caption{$RR_1$ versus $RR$.}\label{figure7}
\end{figure}

There is an important difference between $RR_1$ and $RR$. 
When the outcome's probability is low, the strength of the association would be underestimated with the $RR$. For example, for the case ${(\pi}_{ijE},\pi_{ij\bar{E}},\rho_{ijkE},\rho_{ijk\bar{E}})=(0.1,0.1,0.9,0.1)$, we have that $RR=1$, indicating that there is no association between the outcome and the exposure. However, taking into account the correlation for the individuals who had the event in the previous time, we obtain that ${RR}_1 = 4.8$, which allow us to conclude that an association exists in such a way that the risk for those exposed is almost 5 times the risk for those non-exposed (see Figure \ref{figure7}).

For small values of the outcome's probability of the exposed ($\pi_{ijE}$), we have that $RR<{RR}_1$. In general, the traditional association measures tend to underestimate the association between the outcome and the exposure in individuals who also had the event at a previous time $k$ (see Figure \ref{figure7}).

Furthermore, $RR_1$ reaches the lowest values when the correlation in the non-exposed is high, and also, when the correlation in the exposed is low, being the lowest when $(\rho_{ijkE},\rho_{ijk\bar {E}})= (0.1,0.9)$. The highest values are obtained for high correlations in the exposed and for low correlations in the non-exposed, being the highest when $(\rho_{ijkE},\rho_{ijk\bar{E}}) = (0.9, 0.1)$ (see Figure \ref{figure7}).

\subsection{$RR_0$ versus $RR$}

\begin{figure}[!ht]
  \centering
  \includegraphics [scale=0.6]{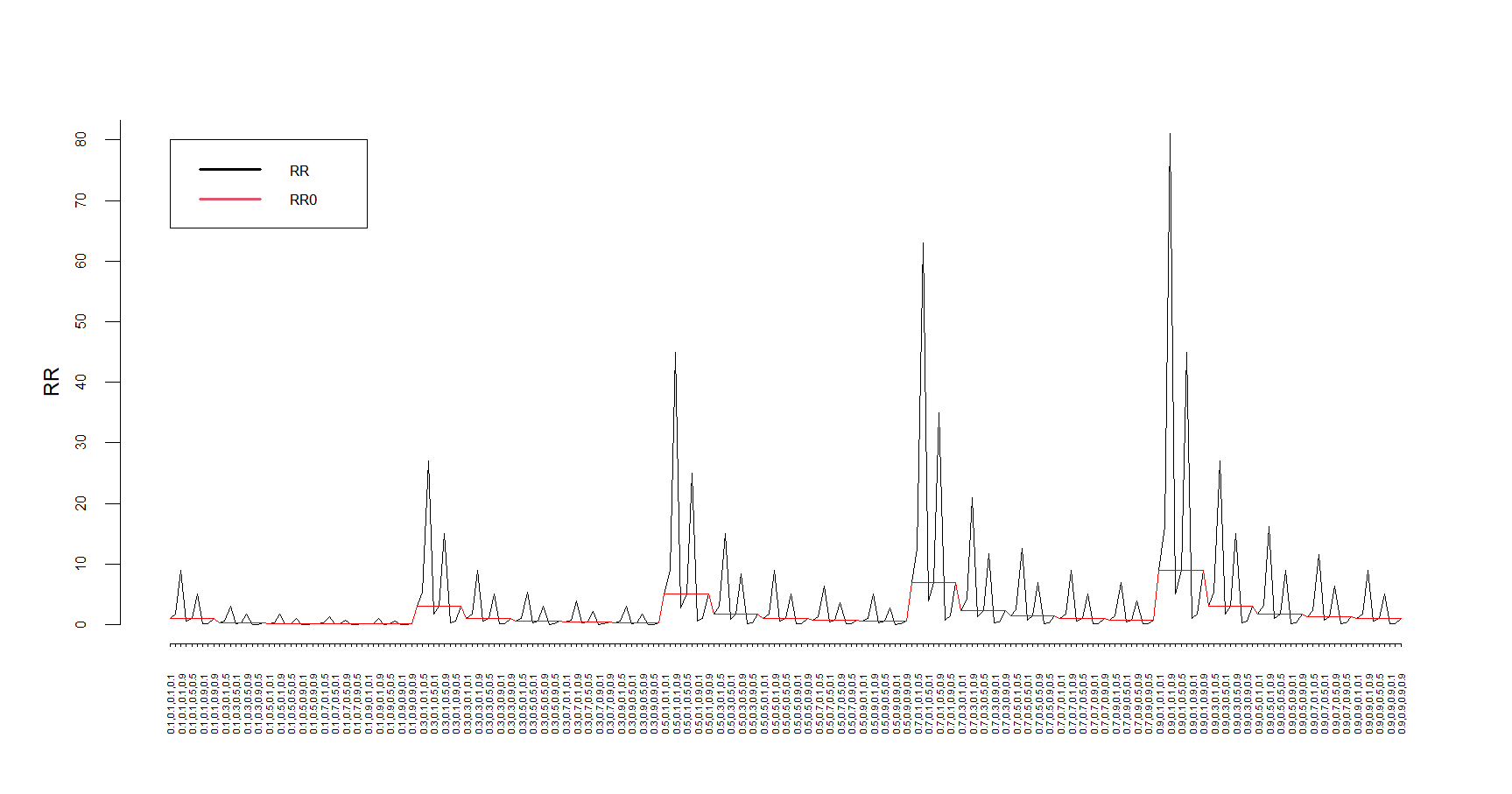}
  \caption{$RR_0$ vs $RR$}\label{figure8}
\end{figure}

In general, $RR$ underestimates the association between the outcome and the exposure for individuals who did not have the event in a previous time $k$. This underestimation becomes more evident as the outcome's probability in the exposed $(\pi_{ijE})$ is greater. For example, for the case $(\pi_{ijE}, \pi_{ij\bar {E}}, \rho_{ijkE}, \rho_{ijk\bar {E}}) = (0.9,0.1, 0.1,0.9)$, we have that $RR = 9$, which indicates that the risk among the exposed is 9 times the risk among the non-exposed. However, taking into account the correlation for the individuals who did not have the event at the previous time, we obtain that ${RR}_0 = 81$, which allow us to conclude that the risk for the exposed is 81 times the risk for the non-exposed. Such a phenomenon occurs because when the outcome's probability in the exposed is high and the correlation is low, the conditional probability is high ($\pi_{Ex | k = 0} = 0.81$). Whereas when the probability of occurrence in the non-exposed is low and the correlation is high, the probability that the event occurs at time $j$ must be very low $(\pi_{\bar {E} j | k = 0} = 0.01)$ (See figure \ref{figure8}).

Furthermore, ${RR}_0$ grows as the outcome's probability in the exposed increases, and then, decreases as the outcome's probability in the non-exposed increases. Nevertheless, the behavior in terms of correlations is totally opposite to that of ${RR}_1$, since it reaches the lowest value for low correlations in the non-exposed and high correlations in the exposed, the lowest being when $(\rho_{ijkE}, \rho_{ijk\bar {E}}) = (0.9,0.1)$, and the highest when the correlation in the exposed is low and in the non-exposed is high, being the highest when $(\rho{ijkE},\rho_{ijk\bar {E}}) = (0.1,0.9)$ (See figure \ref{figure8}).

\section{Application to Contraception Data}\label{sec5}

These data have been analyzed in \cite{Machin88}, \cite{Fitzmaurice04} as well as \cite{Wakefield13}. This dataset comes from a randomized clinical trial, in which 1151 women were randomized to 100 or 150 mg doses of depot medroxyprogesterone acetate (DMPA). This contraceptive was injected on the day of randomization and three more doses at 90-day intervals, the follow-up period was three months after the last injection. The outcome was amenorrhea and was measured in the 4 visits. In this analysis, the data of the women with complete information on the outcome in the 4 visits were included ($n=714$), and patients with doses of 150 mg were considered ``exposed'' and with 100 mg as ``non-exposed''. In general, the probability of amenorrhea increased over time in both exposed and non-exposed (see Table \ref{table3}).
\begin{table}[!ht]
    \caption{\small{Risk of amenorrhea ($\%$) for each visit.}}\label{table3}
\centering {\small
\begin{tabular}{ccccc}\hline
 Dose   & Visit 1 & Visit 2 &  Visit 3  & Visit 4 \\ \hline
 150 mg & $16.1$  & $29.7$  &  $47.9$   & $53.5$  \\
 100 mg & $17.7$  & $25.5$  &  $37.1$   & $51.8$  \\ \hline
\end{tabular}}
\end{table} 

For the second visit, considering all patients, no association was found between the dose and the occurrence of amenorrhea. This conclusion is the same for those who had had amenorrhea on the first visit, but it is the opposite for those who had not had it on the first visit, since in this case, patients with higher dose have a greater risk of amenorrhea. On the third visit, when considering all the patients, an association was found: the higher dose produced a greater risk of amenorrhea, the conclusion was the same for those who did not have the outcome on the second visit. The opposite occurred among those who had amenorrhea on the second visit, since in this case there was no association, that is, the dose did not modify the risk of the outcome. On the fourth visit, something similar to the second occurred, no association was obtained when considering all the patients or those who had not reported amenorrhea on the third visit. While for those who had reported it, an association was found, the higher dose produced a greater risk of occurrence of amenorrhea (see Table \ref{table4}).

In this application, we can see how the conclusions would change if we take into account the correlation with previous times. In general, we would say that the dose does not seem to affect the risk of amenorrhea; however, we see that this actually occurs for women who have had it before, but for those who have not, a higher dose does increase the risk to have the outcome. Therefore, the recommendation would be to be very careful with the dose in those women without a history of amenorrhea (see Table \ref{table4}).
\begin{table}[!ht]
    \caption{\small{Estimation, point and $95\%$ confidence interval, of traditional and conditional relative risks ($\%$) for each visit.}}\label{table4}
\centering {\small
\begin{tabular}{cccc}\hline
                 & Visit 2       &  Visit 3     & Visit 4     \\ \hline
$\hat{RR}$       & $1.17$        &  $1.29$      & $1.07$      \\
$95\%\text{ CI}$ & $0.92-1.48$   &  $1.08,1.53$ & $0.93,1.23$ \\ \hline 
$\hat{RR}_1$     & $0.93$        &  $1.08$      & $0.93$      \\ 
$95\%\text{ CI}$ & $0.70-1.23$   &  $0.93-1.25$ & $0.83-1.03$ \\ \hline
$\hat{RR}_0$     & $1.39$        &  $1.40$      & $1.03$      \\ 
$95\%\text{ CI}$ & $1.01-1.93$   &  $1.07-1.84$ & $0.66-1.39$ \\ \hline
$\hat{\rho}_E$   & $0.28$        &  $0.43$      & $0.58$ \\ 
$\hat{\rho}_{\bar{E}}$ & $0.41$ &  $0.72$      & $0.73$ \\ \hline
\end{tabular}}
\end{table} 

\section{Conclusions}\label{sec6}

In this paper, two brand new approaches to calculate association measures in longitudinal studies are proposed. We based such proposals on the definition of risk, taking into account what happened in previous times. We find that the probability of occurrence of a dichotomous event, conditioned to what happened at the previous moment, involves the correlation between these variables. 

In general, the coverage probability of the proposed confidence intervals for ${RR}_1$ is close to the required value, but when the correlation in the non-exposed is high, this probability takes smaller values. For the proposed confidence intervals for ${RR} _0$, the coverage probability is close to or higher than the predetermined one.

The conclusions about association may be different depending on whether or not the correlation between the times is taken into account, showing the fundamental importance of our proposal in both observational and experimental studies. As an example of an experimental study, we consider an actual application in contraception data, in which the difference between the analysis with the traditional relative risk and our proposals is more than evident, since in visits two and four, it would have been concluded that the dose it is not associated with an increased risk of outcome for all women, $RR=1.17$ $(95\% CI: 0.92-1.48)$ and  $RR=1.07$  $(95\% CI: 0.93-1.23)$, respectively, when in fact, the risk is greater with a dose of DMPA 150 mg for those women who have not previously suffered amenorrhea (see $RR_0$ in Table \ref{table4}).

Future research can be oriented to obtain adjusted measures of association including the correlation with what happened in the previous time, taking into account confounding factors. Similarly, the possibility remains open to study further $RR_1$ and $RR_0$ considering more than one previous measurement. Furthermore, confidence intervals can be developed for $RR_1$ and $RR_0$ when the exposure is not constant over time, and also, when confounding variables are present. Finally, estimating $RR_1$ and $RR_0$ in the context of multilevel modeling is a possibility.

\section*{Statements and Declarations}

The authors declare that they have no known competing financial interests or personal relationships that
could have appeared to influence the work reported in this article.

%\nocite{*}
\bibliography{references}
\bibliographystyle{apalike}

%\appendix

\end{document}